# Do High-Performance Image-to-Image Translation Networks Enable the Discovery of Radiomic Features? Application to MRI Synthesis from Ultrasound in Prostate Cancer

Mohammad R. Salmanpour[1][0000-0002-9515-789], Amin Mousavi[2][0009-0009-0232-3587], Yixi Xu[3][0000-0003-0397-8832], William B. Weeks[3][0000-0002-9918-1360], Ilker Hacihaliloglu[1,4][0000-0003-3232-8193]

[1] Department of Radiology, University of British Columbia, Vancouver, BC, Canada [*m.salmanpour@ubc.ca*], [2] Department of Computer, Abhar Branch, Islamic Azad University, Abhar, Iran, [3] AI for Good Research Lab, Microsoft Corporation, Redmond, WA, US, [4] Department of Medicine, University of British Columbia, Vancouver, BC, Canada

**Abstract.** This study investigates the foundational characteristics of image-to-image translation networks, specifically examining their suitability and transferability within the context of routine clinical environments, despite achieving high levels of performance, as indicated by a Structural Similarity Index (SSIM) exceeding 85%. The evaluation study was conducted using data from 794 patients diagnosed with Prostate cancer (PCa). To synthesize MRI from Ultrasound (US) images, we employed five widely recognized image-to-image translation networks in medical imaging: 2D-Pix2Pix, 2D-CycleGAN, 3D-CycleGAN, 3D-UNET, and 3D-AutoEncoder. For quantitative assessment, we report four prevalent evaluation metrics: Mean Absolute Error (MAE), Mean Square Error (MSE), Structural Similarity Index (SSIM), and Peak Signal to Noise Ratio (PSNR). Moreover, a complementary analysis employing Radiomic features (RF) via Spearman correlation coefficient was conducted to investigate, for the first time, whether networks achieving high performance (SSIM>85%) could identify low-level RFs. The RF analysis showed 75 features out of 186 RFs were discovered via just 2D-Pix2Pix algorithm while half of RFs were lost in the translation process. Finally, a detailed qualitative assessment by five medical doctors indicated a lack of low-level feature discovery in image-to-image translation tasks. This study indicates current image-to-image translation networks, even with a high performance (SSIM>0.85), don't guarantee the discovery of low-level information which is essential for the integration of synthesized MRI data into regular clinical practice.

## 1    Introduction

Ultrasound (US) imaging offers real-time, cost-effective, and bedside diagnostic capabilities. Recent advancements in transducer technology have propelled the adoption of point-of-care ultrasound (POCUS), extending its utility across various clinical domains and beyond hospital settings, particularly in resource-constrained environments [1]. However, persisting challenges including low signal-to-noise ratio (SNR) and the prevalence of imaging artifacts continue to hinder the optimal utilization of both traditional cart-based and POCUS imaging systems [1]. Recently, image-to-image translation methods have been investigated to overcome these problems.

In [2], authors proposed synthesizing pseudo-CT images from US scans.[3] investigated pseudo-anatomical display images generated from ultrasound data. [4] explored a generative attention network for synthesizing X-ray spine images from US scans. Additionally, [5] introduced a self-supervised method for synthesizing MRI fetal brain images from US scans. Quantitative analysis of the synthesized images encompassed the evaluation of key metrics, including MAE, MSE, SSIM, and PSNR [6]. Since these metrics are not always sufficient to capture the complexity of the data and the underlying biological processes [7, 8, 9, 10], certain studies have examined the enhancements achieved in downstream tasks, such as image classification or segmentation [11, 12].

Radiomic features (RF), encompassing spatial distribution, shape, intensity, and texture of radiological structures within the translated images, could offer complementary insights to these metrics. Such analysis could ensure that critical diagnostic information such as changes and characteristics in tissues is not lost in the translation process [13]. However, to date, no study has explored the significance of RF analysis in the context of image synthesis.

Therefore, this study aims to compare RFs extracted from the original high-resolution data with those from translated US images (synthesized MRI images), examining the visual similarity of images at detailed information (low feature) levels. Furthermore, this effort represents one of the initial studies into this area, particularly concerning US data. Specifically, we focus our attention on Prostate cancer (PCa), the second most common cancer in men and the fifth leading cause of cancer-related deaths [14]. We make the following contributions: 1) We investigate the synthesis of MRI-like images from US data, utilizing five widely employed deep learning (DL) methods for medical image synthesis: 2D-CycleGAN, 2D-Pix2pix, 3D-CycleGAN, 3D-AutoEncoder, and 3D-UNET [15]. Notably, this marks the first attempt to synthesize prostate MRI images from US data, with a specific focus on the detection of malignant lesions in PCa. 2) We extend the conventional quantitative analysis by investigating the capability of high-performing networks, achieving SSIM scores exceeding 85%, in identifying low-level RFs. This novel exploration sheds light on the intricate relationship between image quality metrics and the extraction of clinically relevant features, providing valuable insights for future research in medical image analysis. 3) We contribute to the qualitative evaluation domain by engaging five experienced medical professionals in assessing the synthesized MRI images. Their qualitative insights provide a nuanced understanding of the clinical utility and perceptual fidelity of the synthesized images, offering valuable feedback for refining and validating the proposed synthesis methodologies.

## 2    Material and Methods

**Patient Data and Preprocessing Steps.** We employed 794 patients with PCa who had US, T2-weighted MRI, and masks delineated on both images from The Cancer Imaging Archive (dataset name: Prostate-MRI-US-Biopsy) [16]. US scans were performed with Hitachi Hi-Vision 5500 7.5 MHz or the Noblus C41V 2-10 MHz end-fire probe while MR imaging was performed on a 3 Tesla Trio, Verio or Skyra scanner (Siemens, Erlangen, Germany). All US and MRI images were already registered by our clinical collaborators, cropped to 128×128×64 cubic millimeters from the prostate gland center, normalized using the min-max function, and then utilized in training.

**DL-based Image Translation.** Five image-to-image translation algorithms were investigated to synthesize MRI from US images: 2D-CycleGAN, 2D-Pix2Pix, 3D-CycleGAN, 3D-AutoEncoder, and 3D-UNET [15]. The dataset with 794 patients was split into 3 sections including 75% for training, 10% for training validation, and 15% for external testing. The performance of the networks is assessed through 4 evaluation metrics: MAE, MSE, SSIM, and PSNR. We performed 3-fold cross-validation and reported average in all the experiments. Network parameters are listed in the supplemental file. Individual 2D slices from the 3D volumetric data were used as input for the 2D models. The 3D volume was then reconstructed by integrating these 2D slices. Thus, the assessment of all images generated by both 2D, and 3D networks was based on the 3D volumetric data.

**RF Analysis.** Radiomics feature generator within ViSERA (*visera.ca*), extensively standardized in reference to the Image Biomarker Standardization Initiative [17] was utilized to extract a total of 186 standardized RFs, including 2 local intensity (LI), 18 intensity-based statistics (IS), 23 intensity histogram (IH), 7 Intensity-Volume Histogram (IVH), and 136 texture features containing gray level co-occurrence matrix (GLCM; 50 features), gray level run-length matrix (GLRLM; 32 features), gray level size zones (GLSZM; 16 features), gray level distance zone matrix (GLDZM; 16 features), neighborhood gray-tone difference matrix (NGTDM, 5 features), and neighboring gray level dependence matrix (NGLDM; 17 features). RF analysis was conducted using the Spearman correlation function and paired t-test. This analysis encompassed 186 RFs extracted from the segmented prostate gland of both the original and synthetic MRI images. Moreover, we did not utilize any morphological features in this study due to the utilization of identical masks for extracting such characteristics from various images, encompassing original and synthetic MRIs.

**Qualitative Analysis.** In the qualitative validation process, we initially present 15 synthetic MRI images randomly selected from the external testing dataset to five medical doctors with over five years of experience (row 2 in Table 1). Their task is to differentiate between original and synthetic prostate MRI images (row 3 in Table 1). Following this, we provide the medical doctors with specified original and synthetic MRI images and pose eight additional questions to them, prompting them to visually compare and evaluate the synthetic MRI images in relation to the original MRI and US images (rows 4-11 in Table 1). To ensure the reliability and validity of our qualitative analysis questions, we designed them and had them reviewed by an independent professional. The evaluating doctors were blinded to the imaging source (original vs. synthesized MRI) during their assessments. Additionally, we performed an average Inter-Item Correlation statistical test to validate the survey's internal consistency, reliability, and validity.

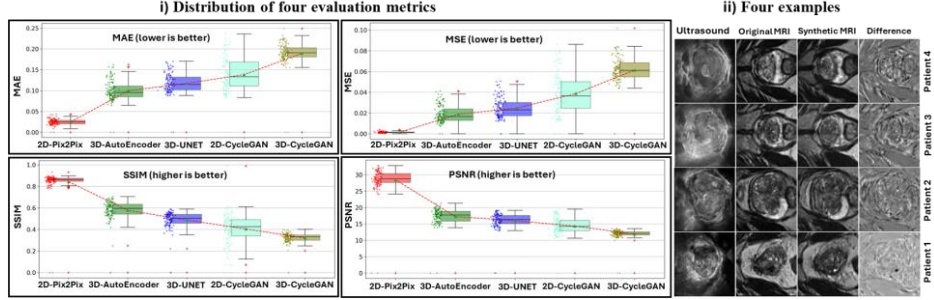

**Fig. 1. (i)** A distribution of four quantitative evaluation metrics: MAE, MSE, SSIM, and PSNR for 2D-Pix2Pix, 2D-CycleGAN, 3D-CycleGAN, 3D-AutoEncoder, and 3D-UNET in synthesizing MRI images from US images, **(ii)** four examples of synthetic MRI images provided by 2D-Pix2Pix. Rows show Ultrasound, Original MRI, Synthetic MRI, difference between original and synthetic MRI images. Columns show different patients. All synthetic images had SSIMs>0.85. **High resolution of this figure is provided in "Code and Data Availability" section.**

## 3      Results and Discussions

**DL-based Image Translation Quantitative Assessment.** As shown in Fig. 1 (i), 2D-Pix2Pix network significantly outperformed the other four networks, with an average MAE of 0.026±0.007, MSE of ~ 0.001±0.001, SSIM of 0.855±0.032, and PSNR of 28.831±2.067 (P-values<0.01, paired t-test, compared to the performance provided from other networks). 2D-CycleGAN had an average MAE of 0.141±0.037, MSE of 0.040±0.018, SSIM of 0.372±0.129, and PSNR of 14.538±2.200, while 3D-CycleGAN had an average MAE of 0.192±0.018, MSE of 0.062±0.011, SSIM of 0.301±0.034, and PSNR of 12.124±0.726. In addition, 3D-UNET and 3D-AutoEncoder provided average MAEs of 0.119±0.021 and 0.102±0.022; MSEs of 0.025±0.009 and 0.019±0.008; SSIMs of 0.466±0.059 and 0.553±0.068; and PSNRs of 16.259±1.578 and 17.499±1.761, respectively. Fig. 1 (ii) displays images generated by 2D-Pix2Pix networks for four test patients, highlighting qualitative results with high SSIM values and demonstrating the closeness between the original and synthesized MRI images.

**RF Analysis.** Koo and Li [18] provided a guideline that categorizes correlation coefficients as follows: i) below 0.50 is poor, ii) 0.50-0.75 is moderate, iii) 0.75- 0.90 is good, and iv) above 0.90 is excellent. Therefore, this research employed a threshold of 0.50 to distinguish between groups. Therefore, in our RF analysis, as shown in Fig. 2., feature similarity amounts enabled us to divide RFs into 3 sub-sections, i) Group 1: the low-level RFs were successfully discovered by synthetic MRI images generated through majority of algorithms, ii) Group 2: the low-level RFs were successfully discovered from synthetic MRI images generated by 2D-Pix2Pix algorithm only, and iii)

Group 3: the low-level RFs extracted from synthetic MRI images were not successfully discovered, even with high performance algorithm 2D-Pix2Pix.

As shown in Fig. 2. (i), most algorithms, even with low performance (SSIM<0.6), enable the generation of synthetic MRI images, leading to the discovery of 18 RFs including 1 IS, 2 NGLDM, 4 GLRLM, 2 GLSZM, 6 GLDZM, and 3 NGTDM features. As depicted in Fig. 2. (ii), the analysis revealed that 75 RFs extracted from synthetic MRI images produced by 2D-Pix2Pix network (demonstrating high performance with SSIM>0.85) exhibited a proportional relationship between the quantitative performance of network and the discovery of low-level features. Thus, Group 2 includes 75 RFs with correlation > 0.5 including 5 IS, 17 IH, 2 IVH, 26 GLCM, 6 NGLDM, 12 GLRLM, 3 GLSZM, 3 GLDZM, and 1 NGTDM. Moreover, Group 3 demonstrated that none of the algorithms, including the one with high performance, facilitated the discovery of 93 low-level features, including 2 LI, 12 IS, 6 IH, 5 IVH, 24 GLCM, 9 NGLDM, 16 GLRLM, 11 GLSZM, 7 GLDZM, 1 NGTDM features (see Fig. 2. (iii)).

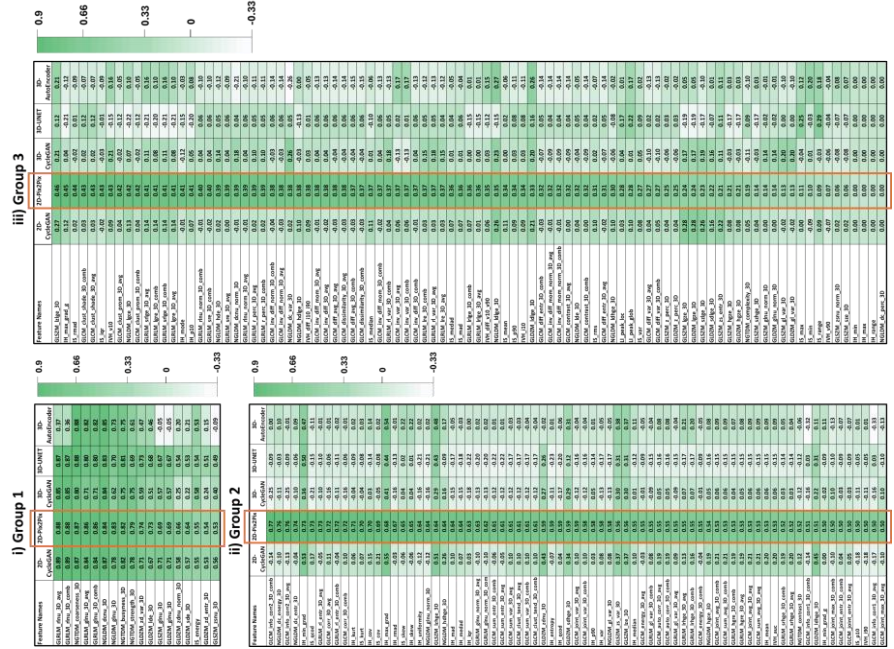

**Fig. 2.** Different Radiomic feature (RF) groups provided by RF Analysis. i) Group 1 showed 18 low-level RFs successfully discovered by synthetic MRI images generated through majority of algorithms, ii) Group 2 showed 75 low-level RFs were successfully discovered from synthetic MRI images generated by 2D-Pix2Pix, and iii) Group 3 showed synthetic MRI images generated by the current generative networks couldn't discovered 93 low-level RFs. **High resolution of this figure is provided in "Code and Data Availability" section.**

**Qualitative Analysis.** The qualitative evaluation of synthetic MRI images, guided by feedback from experienced medical practitioners, was centered on assessing the fidelity of anatomical delineation and tissue contrast compared to original MRI counterparts. Five doctors assessed the perceptual clarity of anatomical structures, precision of boundary delineation, and the presence of any discernible artifacts, crucial factors underpinning the integration of synthetic MRI into clinical practice. Doctors' evaluations were highly consistent, with an average Inter-Item Correlation Coefficient of 0.99, affirming the survey's reliability. As mentioned previously, each doctor possessed extensive expertise exceeding five years in the interpretation of diverse medical imaging modalities including MRI and US (Table 1, row 2). All five experts could discriminate synthetic MRI from original MRI (row 3). Despite average SSIM > 0.85, all doctors believed that the quality of synthetic MRI images (in terms of detailed information) was not comparable with the original ones (row 4).

Furthermore, practitioners remarked upon the salient presence of artifacts with-in synthetic MRI images, serving as a distinguishing hallmark vis-à-vis their authentic counterparts (row 5). All experts believed that diagnosis using the synthetic MRI (even with SSIM>0.85) is difficult, compared to original MRI images (row 6). Although all doctors believed that synthetic MRI images added no value compared to the original MRI (row 7), some experts expressed that synthetic MRI images added value to the diagnosis process compared to US images (row 8) in terms of anatomical structure. Nonetheless, the majority consensus underscored the perceptible discrepancies in resolution and contrast levels between synthetic and original MRI images (row 9). Collectively, practitioners underscored the lack of detailed anatomical information, particularly pertaining to low-level features, within synthetic MRI images - a pivotal focus of this inquiry. Thereby, they were convinced that there are potential clinical benefits of using synthetic MRI images and strongly support the integration of synthetic MRI technology into regular clinical practice if the synthetic MRI images include detailed information (rows 10 and 11). One notable benefit is that synthetic MRIs can improve diagnosis in emergency situations where access to MRI scans is limited.

**Table 1.** Qualitative analysis of synthetic MRI by 5 medical doctors (D). **High resolution of this figure is provided in "Code and Data Availability" section.**

| Questions (Q), Scoring system: 0= zero, 1= low, 2=intermediate, 3=high, 4=very high | D 1 | D 2 | D 3 | D 4 | D 5 |
|---|---|---|---|---|---|
| Q1: What is your medical specialty and how many years of experience do you have in interpreting MRI and ultrasound images? (years) | >5 | >6 | >5 | >5 | >5 |
| Q2: How many doctors could discriminate the synthetic MRI from the original MRI properly? (15 external testing images existed) | 15 | 15 | 15 | 15 | 15 |
| Q3: After specifying synthetic and original MRI for you, how would you rate the overall quality of synthetic MRI images compared to original MRI? | 1 | 2 | 1 | 1 | 1 |
| Q4: Are there any noticeable artifacts or inaccuracies in the synthetic MRI images? | 4 | 2 | 4 | 4 | 3 |
| Q5: How confident are you in making a diagnosis based on synthetic MRI images versus original MRI? | 1 | 1 | 1 | 1 | 1 |
| Q6: Do synthetic MRI images offer any additional diagnostic information compared to the original MRI images? How much? | 0 | 0 | 0 | 0 | 0 |
| Q7: Do synthetic MRI images offer any additional diagnostic information compared to the original Ultrasound images? How much? | 2 | 2 | 3 | 2 | 3 |
| Q8: How do you assess the resolution and contrast of the synthetic MRI images, compared to original MRI images? | 1 | 2 | 1 | 2 | 2 |
| Q9: In your opinion, how much are the potential clinical benefits of using synthetic MRI images? | 4 | 3 | 3 | 4 | 3 |
| Q10: Would you support the integration of synthetic MRI technology into regular clinical practice? How much? | 4 | 4 | 4 | 4 | 4 |

**Discussions.** We have shown that RF analysis is vital to address the limitations of standard metrics such as MAE, MSE, SSIM, and PSNR in image-to-image translation tasks. RFs extend beyond above metrics, offering a more comprehensive analysis that identifies important characteristics such as shape, intensity, texture, and patterns that are used for diagnostics. These features are pivotal for unraveling the underlying biological and pathological information, providing a richer understanding of the intricacies within the synthesized medical images [13]. Our analysis investigates that translated images retain clinically relevant information, bridging the gap between statistical accuracy and clinical utility, which is often overlooked by conventional evaluation metrics.

We believe that this deeper level of analysis is essential in a clinical context, as it can reveal subtle changes and characteristics in tissues that might be crucial for accurate diagnosis, disease monitoring, and treatment planning [13]. By integrating RFs, we ensure that the image translation networks are not only statistically accurate but also effective and meaningful in real-world medical applications, enhancing the reliability and utility of these technologies in healthcare settings [19].

RF analysis in this study indicated that RFs can be divided into three sections in US to MRI translation, including i) a set of low-level features (RFs) that can be discovered by the majority of networks, even with low-performance algorithms, ii) a set of low-level features that can be discovered by high-performances networks only, and iii) a set of low-level features currently undetectable by any existing networks (high or low performance networks). RF analysis, beyond conventional metrics quantifying overall error and similarity, obviously shows that current translation networks, even 2D-Pix2Pix with SSIM>85% are not able to discover half of RFs (93 out of 186 RFs in Group 3). Group 1 showed that 18 RFs out of 186 features can be discovered by the majority of algorithms, even with low-performance algorithms while Group 2 indicated that just 75 RFs can be roughly restored by high-performance translation networks (2D-Pix2Pix).

Thus, the identification of RFs is contingent upon the efficacy of algorithms such as MAE, MSE, SSIM, and PSNR. Notably, algorithms demonstrating superior performance, exemplified by 2D-Pix2Pix, facilitate the synthesis of MRI images, thereby enabling the discernment of certain low-level RFs. Despite the marked significant enhancement observed in the similarity index of RFs derived from synthetic MRI images generated by 2D-Pix2Pix in comparison to other image-to-image algorithms across various groups (P-Value<0.05, paired t-test), there persists a pressing imperative to refine image-to-image translation networks to optimize the performance of low-level features.

Qualitative analysis indicated that differences in quality, existing artifacts, resolution, and contrast between synthetic and original MRI images enabled all doctors to successfully discriminate between the synthetic and original MRI images. Furthermore, all doctors expressed that diagnosis using the synthetic MRI (even with SSIM>0.85), compared to original MRI images, is difficult and the synthetic MRI images didn't cover all low-level features which are essential for successful diagnosis. Thus, it can be concluded that translation networks with high performances didn't guarantee appropriate discovery of low-level features. However, some doctors expressed that the synthetic MRI images added value to the diagnosis process beside US images. They believed that enhancing the detailed information in synthetic MRI images is necessary for integrating synthetic MRI technology into regular clinical practice. This is especially important in low-resource settings where access to traditional MRI is challenging, as pursued in this study.

Our RF analysis indicated that 74% of IH features belonged to Group 2, which highly depends on network performance. These features offer a comprehensive quantitative analysis of tumor traits in medical images, improving cancer diagnosis and assessment. By quantifying pixel intensities, they reveal details about tumor heterogeneity, the microenvironment, and responses to treatment, indicating cellular complexities. Statistics like Mean, Variance, Skewness, and Kurtosis analyze the distribution's tendency and shape, while Median, Mode, and Percentiles examine the data's central aspects and variability. Additionally, metrics such as Entropy, Uniformity, and Gradient evaluate image texture and edges, crucial for assessing tissue characteristics and aiding diagnosis.

GLCM features, complementing visual assessment, are a statistical method to analyze image texture. In this study, 53% of GLCM features belonged to Group 2 while the remaining 47% of features went to Group 3, indicating disabilities of the current networks to discover these kinds of features. These features quantify the co-occurrence of gray levels at specific offsets, providing information about homogeneity, contrast, and other textural properties of images. Radiologists use GLCM features to differentiate between healthy and abnormal tissues, aiding in disease detection and prognosis.

Most NGTDM features fell into Group 1, showing a general consistency with network performance. These features examine the local differences in gray-tone intensities in medical images, offering crucial texture information that improves the accuracy of visual assessments. This assists in differentiating between malignant and benign lesions, thereby enhancing diagnostic decisions. All features of LI and most features of IS fall under Group 3. These characteristics help measure minor changes in the intensity of individual pixels in tumor areas, enhancing visual evaluations. Furthermore, IVH features, indicating the relationship between a gray level area and the fraction of the volume of the histogram, offer a comprehensive perspective on the distribution of intensity, assisting medical professionals in grasping the diversity within tumors. Further, the majority of GLRLM, GLSZM, NGLDM, and GLDZM belonged to Group 3. These features provide insights into tumor heterogeneity, spatial patterns, and microstructure, guiding clinical decisions and patient management.

A recent study [20] has developed a novel technique for enhancing MRI-to-CT image conversion, utilizing a loss function derived from GLCM to reproduce texture features more accurately in generated CT images. This method surpasses traditional pixel-based approaches by focusing on improving texture quality, potentially offering significant benefits to medical image synthesis and its clinical applications. Considering that RFs derived from multiple imaging techniques provide extra information for various purposes, the selection of the most pertinent features for inclusion in the loss function grows more complicated and will be left for a future study. Additionally, we plan to extend our research to larger datasets and other medical imaging datasets. We also aim to explore newer architectures, such as diffusion models, to broaden the applicability of our findings in future studies.

## 4      Conclusions

In summary, this study, employing RF and qualitative analysis conducted by five experienced physicians, determined that synthetic MRI images produced by existing image-to-image translation algorithms, despite achieving a high SSIM performance > 0.85, fail to facilitate the detection of crucial low-level features essential for accurate diagnosis and decision-making processes. Consequently, there is a pressing need for further refinement of image-to-image translation networks to enhance the performance of low-level features. Future work will involve exploring diffusion model-based analysis and developing loss functions that integrate radiomic features (RF) into their design, aiming to enhance the rate of feature discovery.

**Acknowledgments.** This work was supported by the Mitacs Accelerate program (AWD-024298-IT33280), the NSERC Discovery Grant (RGPIN-2023-03575), The Canadian Foundation for Innovation (CFI)- John R. Evans Leaders Fund (JELF) program grant number 42816, and Microsoft's AI for Good Lab. **Disclosure of Interests.** The authors have no competing interests to declare that are relevant to the content of this article.
**Code and Data Availability.** All relevant information: RF names, codes, high resolution figures and tables are available on: *https://github.com/MohammadRSalmanpour/MRI-to-US-Translation_MICCAI/*